%% file: MajkDaCosta.tex
\documentclass{llncs}
\usepackage[german,english]{babel}
\usepackage{times,theorem}

\usepackage{latexsym,color,graphics}
\usepackage{url}
\usepackage{epsfig}
\usepackage{amssymb}
\usepackage{amsmath}
\usepackage{amsfonts}

\input{macros-general1}

\input{macros-l1}
\title{On Paraconsistent Weakening of Intuitionistic Negation}
\author{Zoran Majki\'c}
\institute{International Society for Research in Science and Technology \\
PO Box 2464 Tallahassee, FL 32316 - 2464 USA \\
\email{majk.1234@yahoo.com} }

\newtheorem{theo}{Theorem}
\newtheorem{propo}{Proposition}
\newtheorem{coro}{Corollary}

\begin{document}

\maketitle

\begin{abstract}
In \cite{Majk08dc}, systems of weakening of intuitionistic negation
logic called Z$_n$ and CZ$_n$ were developed in the spirit of da
Costa's approach(c.f. \cite{Costa74}) by preserving, differently
from da Costa, its fundamental properties: antitonicity, inversion
and additivity for distributive lattices.
 However, according to \cite{OmWa10}, those systems turned out to be not paraconsistent but extensions of intuitionistic logic.
 Taking into account of this result, we shall here make some observations on the modified systems of Z$_n$ and
 CZ$_n$, that are paraconsistent as well.
\end{abstract}

\section{Introduction}
 The big challenge for paraconsistent logics is to avoid
 allowing contradictory theories to explode and derive anything else
 and still to reserve a respectable logic, that is, a logic capable of drawing reasonable
 conclusions from  contradictory theories.\\
   There are different approaches to paraconsistent logics: The first is the non constructive approach, based on abstract logic
 (as LFI \cite{CaCM06}), where  logic connectives and their
 particular semantics are not considered. The second is the constructive approach and is divided in
 two parts: axiomatic proof theoretic (cases of da Costa \cite{Costa74} and
 \cite{AnBe75,Bate80,Bate00}),  and many-valued
 (case \cite{Majk06ml}) model theoretic based on truth-functional valuations (that
 is, it satisfies the truth-compositionality principle). The best
 case is when we obtain both proof and model theoretic definition which are mutually sound and
 complete.\\
 One of the main founders with Stanislav Jaskowski \cite{Jask48}, da
Costa,  built his propositional paraconsistent system $C_\omega$ in
\cite{Costa74} by weakening the logic negation operator $\neg$, in
order to avoid the explosive inconsistency \cite{CaCM06,CaMa02} of
the classic propositional logic, where  the ex falso quodlibet proof
rule $~~\frac{A, ~~\neg A}{B}$ is valid. In fact, in order to avoid
this classic logic rule, he changed the semantics for the negation
operator, so that:
\begin{itemize}
  \item NdC1:  in these calculi the principle of non-contradiction,
  in the form $\neg(A \wedge \neg A)$, should not be a generally valid
  schema, but if it does hold for  formula $A$, it is a well-behaved
  formula, and is denoted by $A^\circ$;
  \item NdC2: from two contradictory formulae, $A$ and $\neg A$, it
  would not in general be possible to deduce an arbitrary formula
  $B$. That is it does not hold the falso
quodlibet proof rule $~~\frac{A, ~~\neg A}{B}$;
  \item NdC3: it should be simple to extend these calculi to corresponding
  predicate calculi (with or without equality);
  \item NdC4: they should contain  most parts of the schemata and
  rules of  classical propositional calculus which do not infere
  with the first conditions
\end{itemize}
In fact Da Costa's paraconsistent propositional logic is made up of
the unique Modus Ponens inferential rule (MP), $A, A \Rightarrow B
~\vdash ~B$, and two axiom subsets. But before stating them we need
the following definition as it is done in da Costa's systems(c.f.
\cite[p.500]{Costa74}), which uses three binary connectives,
$\wedge$ for conjunction, $\vee$ for disjunction and $\Rightarrow$
for implication:

\begin{definition}\label{defn:daCosta(n)} Let $A$ be a formula and $1 \leq n < \omega$.
Then, we define $A^\circ, A^{\text{n}}, A^{\text{(n)}}$ as
follows:\\
\noindent \ \ \ $A^\circ  =_{\text{def}} \neg (A \land \neg A)$,
\noindent \ \ \ $A^{\text{n}} =_{\text{def}}
A^{\overbrace{\circ\!\circ\!\cdots\!\circ}^n}$, and \noindent \ \ \
$A^{\text{(n)}} =_{\text{def}} A^1 \land A^2 \land \dots \land
A^{\text{n}}$.
\end{definition}
The first one is for the positive propositional logic (without
negation), composed by the following eight axioms, borrowed from the
 classic propositional logic of the
Kleene $L_4$ system, and also from the more general propositional
\emph{intuitionistic} system (these two systems differ only
regarding axioms with the negation operator),

\textsc{(IPC$^+$) Positive  Logic Axioms:}\\
 (1) $~~A\Rightarrow (B \Rightarrow A)$\\
 (2)  $~~(A \Rightarrow B) \Rightarrow ((A \Rightarrow (B\Rightarrow  C)) \Rightarrow (A \Rightarrow C))$\\
 (3)  $~~A \Rightarrow (B \Rightarrow (A \wedge B))  $\\
 (4)  $~~(A \wedge B) \Rightarrow A  $\\
 (5)  $~~(A \wedge B) \Rightarrow B  $\\
 (6)  $~~A \Rightarrow (A \vee B)  $\\
 (7)  $~~B \Rightarrow (A \vee B)  $\\
 (8)  $~~(A \Rightarrow C) \Rightarrow ((B \Rightarrow C) \Rightarrow((A \vee B) \Rightarrow C))
 $\\
 and change the original axioms for negation of the classic
 propositional logic, by defining semantics of negation by the
 following subset of axioms:\\
\textsc{(NLA)  Logic Axioms for Negation:}\\
(9) $~~A \vee \neg A$\\
(10) $~~\neg \neg A \Rightarrow A$\\
  (11) $~~B^{(n)} \Rightarrow ((A \Rightarrow B) \Rightarrow (( A \Rightarrow \neg B) \Rightarrow \neg A))~~~~$
(Reductio relativization axiom)\\
 (12) $~~(A^{(n)} \wedge B^{(n)}) \Rightarrow ((A \wedge B)^{(n)} \wedge (A \vee B)^{(n)} \wedge (A \Rightarrow B)^{(n)})$
 \\ $\square$\\
 It is easy to see that the axiom (11) relativizes the classic \emph{reductio} axiom $(A \Rightarrow B) \Rightarrow (( A \Rightarrow \neg B) \Rightarrow \neg
 A)$ (which is equivalent to the contraposition axiom $(A \Rightarrow \neg B) \Rightarrow (B \Rightarrow \neg A)$ and the trivialization
 axiom $\neg(A \Rightarrow A) \Rightarrow B$), \emph{only} for propositions $B$ such
 that $~B^{(n)}$ is valid, and in this way avoids the validity of the
 classic ex falso quodlibet proof rule.  It provides a qualified form of reductio, helping to prevent general validity  of $~B^{(n)}$
 in the paraconsistent logic $C_n$. The axiom (12) regulates
 only the propagation of n-consistency. It is easy to verify that
 n-consistency also propagates through negation, that is, $~~A^{(n)}  \Rightarrow (\neg A
 )^{(n)}$ is provable in $C_n$. So that for any fixed $n$ (from $0$ to $\omega$) we
 obtain a particular da Costa paraconsistent logic $C_n$. \\
 One may regard $C_{\omega}$ as a kind of syntactic limit \cite{CaMa99} of the
 calculi in the hierarchy.\\
 Each $C_n$ is strictly weaker than any of its predecessors, i.e.,
 denoting by $Th(S)$ the set of theorems of  calculus $S$, we
 have:\\
 $Th(CPL) \supset Th(C_1) \supset ... \supset Th(C_n) \supset...\supset
 Th(C_\omega)$.\\
 Thus we are fundamentally interested in the $C_1$ system which is a paraconsistent logic
 closer  to the CPL (Classic propositional logic), that is, $C_1$ is the paraconsistent
 logic of da Costa's hierarchy
 obtained by minimal change of CPL.\\
 It is well known that the classic propositional logic based on the classic 2-valued complete distributive lattice $(\textbf{2},
 \leq)$ with  the set $\textbf{2} = \{0, 1\}$ of truth values, has a
 truth-compositional model theoretic semantics.
 For this da Costa calculi is not given any truth-compositional
 model  theoretic semantics instead.\\
  The \emph{non-truth-functional} bivaluations (mappings from the set of well-formed formulae of
 $C_n$ into the set $\textbf{2}$) used in \cite{CoAl77,LoAl80}
 induce the decision procedure for $C_n$ known as quasi-matrices
 instead. In this method, a negated formulae within truth-tables
 must branch: if $A$ takes the value 0 then $\neg A$ takes the value
 1 (as usual), but if $A$ takes the value 1 then $\neg A$ can take
 either the value 0 or the value 1; both possibilities must be
 considered, as well as the other axioms governing the
 bivaluations.\\
  Consequently, the da Costa system still
 needs a kind of (relative) compositional model-theoretic
 semantics.
 Based on these observations, in \cite{Majk08dc}   are explained some weak properties of Da Costa
 weakening for a negation operator, and was shown that it is not antitonic, differently from the negations in the classic
  and intuitionistic propositional logics (that
 have the truth-compositional
 model  theoretic semantics).
 The axioms for negation in CPL are as follows:\\
 \textsc{(NCLA) Classic Axioms for Negation:} \\
(9) $~~A \vee \neg A$\\
(10c) $~~(A\Rightarrow B) \Rightarrow ((A \Rightarrow \neg B) \Rightarrow \neg A)$\\
(11c) $~~A \Rightarrow (\neg A \Rightarrow B)$\\
(12c) $~~0 \Rightarrow A$, $~~A \Rightarrow 1$ \\
while for the intuitionistic logic we eliminate the
axiom (9).\\
 The negation in the classic and intuitionistic logics are not paraconsistent
 (see for example Proposition 30, pp 118, in \cite{Majk06ml}), so
 that  Majki\'c's idea in \cite{Majk08dc} was to make a weakening
 of the intuitionistic negation by considering only its general
 antitonic property: in fact the formula $(A\Rightarrow B) \Rightarrow (\neg
 B \Rightarrow \neg A)$ is a thesis in both classic and intuitionistic
 logics. Consequently, his idea was to make  da Costa weakening of the \emph{intuitionistic} negation \cite{Majk08dc}, that is, to define the system $Z_n$
for each $n$ by adding the following axioms to the system IPC$^+$:\\\\
\noindent (11) \ \ \ $B^{\text{(n)}} \Rightarrow ((A \Rightarrow B)
\Rightarrow
((A \Rightarrow \neg B) \Rightarrow \neg A))$\\
\noindent (12) \ \ \ $(A^{\text{(n)}} \land B^{\text{(n)}})
\Rightarrow ((A \land B)^{\text{(n)}} \land (A \lor B)^{\text{(n)}}
\land (A
\Rightarrow B)^{\text{(n)}})$\\
\noindent (9b) \ \ \ $(A \Rightarrow B) \Rightarrow (\neg B
\Rightarrow \neg A)$\\
\noindent (10b) \ \ \ $1 \Rightarrow \neg 0$, $\neg 1 \Rightarrow 0$\\
\noindent (11b) \ \ \ $A \Rightarrow 1$, $0 \Rightarrow A$\\
\noindent (12b) \ \ \ $(\neg A \land \neg B) \Rightarrow \neg (A
\lor B)$

\noindent Finally, the hierarchy CZ$_n$ is obtained by adding the
following axiom:

\vskip3pt

\noindent (13b) \ \ \ $\neg (A \land B) \Rightarrow (\neg A \lor
\neg B)$\\\\
The result provided in \cite{OmWa10} is that in the above
formulation of the system Z$_n$, axioms (11), (12) and (12b) are
redundant in the sense that those formulas can be derived from the
other axioms (9b), (10b) and (11b) in addition to IPC$^+$.
Obviously, the formulation of CZ$_n$ is given by adding the axiom
(13b). As a result, systems Z$_n$ and CZ$_n$ do not form a hierarchy
but are single systems. It is also proved that formulas
`$(A\Rightarrow (A\land \neg A))\Rightarrow \neg A$' and
`$A\Rightarrow (\neg A\Rightarrow B)$' can be proved in Z$_n$ which
shows that Z$_n$ and CZ$_n$ are extensions of intuitionistic
propositional calculus and therefore not
paraconsistent.\\
In fact, the introduction of the axiom $\neg 1 \Rightarrow 0$ in the
system Z$_n$ is \emph{not necessary} for the all obtained results in
\cite{Majk08dc}: this formula was responsible for the fact that
Z$_n$ is not paraconsistent.\\
In what follows we will present the properties of this modified
system, by eliminating this formulae from the system $Z_n$.
\section{Paraconsistent weakening of negation}
%
 In what follows we
consider modified systems of $Z_n$ and $CZ_n$ which can be obtained
by eliminating the formula `$\neg 1 \Rightarrow 0$' of axiom (10b)
from the systems $Z_n$ and $CZ_n$. Notice that this axiom is not
necessary in order to have additive modal negation operator that can
be modeled by Birkhoff's polarity as required in \cite{Majk08dc}. We
shall refer to these systems as $mZ_n$ and $mCZ_n$ respectively and
also refer to the modified
axiom as (10b)'.\\
Thus, all results obtained in \cite{Majk08dc} are preserved for this
logic: what we need is only to eliminate the sequent $~ \neg 1
\vdash 0~$ from (5a) in Definition 7 (Gentzen-like system) in
\cite{Majk08dc} as well.\\
Consequently, these modified systems $mZ_n$ and $mCZ_n$ have the
Kripke possible world semantics
 for these two paraconsistent logics (defined by Definition 6 in \cite{Majk08dc}), and based on it, the
 many-valued semantics based on functional hereditary distributive
 lattice of algebraic truth-values. Finally,  this many-valued (and Kripke) semantics, based on model-theoretic entailment, is
 adequate, that is, sound and complete w.r.t. the proof-theoretic da
 Costa axiomatic systems of these two paraconsistent logics $mZ_n$ and
 $mCZ_n$.\\
 We now prove some another results on $mZ_n$ and $mCZ_n$:
\begin{propo}
Following formulas are derivable in $mZ_n$ (we denote by $A \equiv
B$ the formulae $(A \Rightarrow B) \wedge (B\Rightarrow A)$):
\begin{gather}
((A\Rightarrow B) \wedge (A \Rightarrow C))\Rightarrow (A
\Rightarrow (B\wedge C)) \tag{T0} \label{Composition}\\
(A\Rightarrow (B\Rightarrow C))\Rightarrow (B\Rightarrow (A\Rightarrow C)) \tag{T1} \label{Exchange} \\
(A\Rightarrow B)\Rightarrow ((B\Rightarrow C)\Rightarrow (A\Rightarrow C)) \tag{T2} \label{Syllogism} \\
(A\Rightarrow (B\Rightarrow C))\equiv ((A\land B)\Rightarrow C)
\tag{T3} \label{impland}
\end{gather}
\end{propo}
 This is obvious since $mZ_n$ contains IPC$^+$.
\begin{theo}
Systems $mZ_n$ and $mCZ_n$ are paraconsistent.
\end{theo}
\textbf{Proof:}  Just interpret the negation as a function always
giving truth value 1 whereas other connectives interpreted in a
standard way done in two valued for classical propositional
calculus.\\$\square$\\
It should be noted that even though we have the above theorem, the
following formula  $(A \land \neg A) \Rightarrow \neg B $ is still
derivable, as we can show by the following lemma:
\begin{lemma}
The following formulae are derivable in $mZ_n$:
\[ (A \land \neg A) \Rightarrow \neg B \tag*{NEFQ} \label{NEFQ} \]
\[ \neg \neg (A\land B) \Rightarrow (\neg \neg A\land \neg \neg B) \tag{$\heartsuit $} \label{keyformula2}
\]
\[ \neg ((A\ast B)^n)\Rightarrow (\neg (A^n) \lor \neg (B^n)) \tag{$\clubsuit $} \label{keyformula} \]
where $\ast\in \{ \Rightarrow , \land , \lor \}$.
\end{lemma}
\textbf{Proof:} Let us derive NEFQ:

1 \ \ \ $A\Rightarrow (B\Rightarrow A)$ \hfill[(1)]

2 \ \ \ $(B\Rightarrow A)\Rightarrow (\neg A \Rightarrow \neg B)$
\hfill[(9b)]

3 \ \ \ $A\Rightarrow (\neg A \Rightarrow \neg B)$ \hfill[1, 2,
\eqref{Syllogism}]

4 \ \ \ $(A \land \neg A) \Rightarrow \neg B$ \hfill[3,
\eqref{impland}, (MP)]

\noindent Notice that \eqref{NEFQ} is not desirable for some
paraconsistent\\
 Let us derive $\heartsuit $ now. We will only prove the following, since
the case in which $\neg \neg A$ is replaced by $\neg \neg B$ can be
proved analogously:
\[ \neg \neg (A\land B) \Rightarrow \neg \neg A \]
This can be proved easily by making use of axioms (3) and (9b).\\
Let us derive $\clubsuit $ now. The proof runs as follows:

1 \ \ \ $\neg ((A\ast B)^n)\equiv \neg \neg ((A\ast B)^{n-1}\land
\neg (A\ast B)^{n-1})$ \hfill[Definition of $A^n$]

2 \ \ \ $\neg \neg ((A\ast B)^{n-1}\land \neg (A\ast B)^{n-1})
\Rightarrow (\neg \neg (A\ast B)^{n-1}\land \neg \neg \neg (A\ast
B)^{n-1})$ \hfill[\eqref{keyformula2}]

3 \ \ \ $(\neg \neg (A\ast B)^{n-1}\land \neg \neg \neg (A\ast
B)^{n-1})\Rightarrow \neg (A^n)$ \hfill[\eqref{NEFQ}]

4 \ \ \ $\neg (A^n)\Rightarrow (\neg (A^n)\lor \neg (B^n))$
\hfill[(6)]

5 \ \ \ $\neg ((A\ast B)^n)\Rightarrow (\neg (A^n) \lor \neg (B^n))$
\hfill[1, 2, 3, 4, \eqref{Syllogism}, (MP)]

\noindent This completes the proof.
\\$\square$\\
Let us  show now that the axioms (11) and (12) are redundant in the
System $mZ_n$.
\begin{theo}\label{thm:axiom11redundant}
The axioms (11) and (12) are redundant in $mZ_n$ in the sense that
they can be proved by another axioms.
\end{theo}
\textbf{Proof:} The redundance of the axiom (11) can be proved as
follows:

1 \ \ \ $(A \Rightarrow (B\land \neg B))\Rightarrow (\neg (B\land
\neg B)\Rightarrow \neg A)$ \hfill[(9b)]

2 \ \ \ $\neg (B\land \neg B) \Rightarrow ((A \Rightarrow (B\land
\neg B))\Rightarrow \neg A)$ \hfill[1, \eqref{Exchange}, (MP)]

3 \ \ \ $B^{\text{(n)}} \Rightarrow B^1$ \hfill[Definition of
$B^{\text{(n)}}$]

4 \ \ \ $B^{\text{(n)}} \Rightarrow \neg (B\land \neg B)$
\hfill[Definition of $B^1$]

5 \ \ \ $B^{\text{(n)}} \Rightarrow ((A \Rightarrow (B\land \neg
B))\Rightarrow \neg A)$ \hfill[2, 4, \eqref{Syllogism}, (MP)]

\noindent Let us prove the redundance of the axiom (12).
 It would be sufficient to prove the following in
order to prove the desired result:
\[ (A^{\text{(n)}} \land B^{\text{(n)}}) \Rightarrow (A \ast B)^n \tag{$\diamondsuit $} \label{axiom12} \]
Indeed, if we have \eqref{axiom12} at hand then we can prove
\[ (A^{\text{(n)}} \land B^{\text{(n)}}) \Rightarrow (A \ast B)^m \]
for any $1\leq m\leq n$ and combining all these cases, we obtain
\[ (A^{\text{(n)}} \land B^{\text{(n)}}) \Rightarrow (A \ast B)^{\text{(n)}} \]
which is axiom (12). So, we now prove \eqref{axiom12} which runs as
follows:

1 \ \ \ $(A^{\text{(n)}} \land B^{\text{(n)}}) \Rightarrow (((A\ast
B)^{n-1}\land \neg (A\ast B)^{n-1})\Rightarrow (\neg (A^{n-1})\lor
\neg (B^{n-1}))$ \hfill[\eqref{keyformula}]

2 \ \ \ $(A^{\text{(n)}} \land B^{\text{(n)}}) \Rightarrow (((A\ast
B)^{n-1}\land \neg (A\ast B)^{n-1})\Rightarrow ((A^{n-1} \land \neg
(A^{n-1}))\lor (B^{n-1}\land \neg (B^{n-1}))))$ \\ \hfill[1,
Definition\ref{defn:daCosta(n)}]

3 \ \ \ $(((A\ast B)^{n-1}\land \neg (A\ast B)^{n-1})\Rightarrow
((A^{n-1} \land \neg (A^{n-1}))\lor (B^{n-1}\land \neg
(B^{n-1}))))\Rightarrow (\neg ((A^{n-1} \land \neg (A^{n-1}))\lor
(B^{n-1}\land \neg (B^{n-1})))\Rightarrow \neg ((A\ast B)^{n-1}\land
\neg (A\ast B)^{n-1}))$ \hfill[(9b)]

4 \ \ \ $(A^{\text{(n)}} \land B^{\text{(n)}}) \Rightarrow (\neg
((A^{n-1} \land \neg (A^{n-1}))\lor (B^{n-1}\land \neg
(B^{n-1})))\Rightarrow \neg ((A\ast B)^{n-1}\land \neg (A\ast
B)^{n-1}))$ \\ \hfill[2, 3, \eqref{Syllogism}, (MP)]

5 \ \ \ $(\neg (A^{n-1} \land \neg (A^{n-1}))\land \neg(B^{n-1}\land
\neg (B^{n-1}))) \Rightarrow \neg ((A^{n-1} \land \neg
(A^{n-1}))\lor (B^{n-1}\land \neg (B^{n-1})))$ \hfill[(12b)]

6 \ \ \ $(A^{\text{(n)}} \land B^{\text{(n)}}) \Rightarrow
((A^n\land B^n)\Rightarrow (A\ast B)^n)$ \hfill[4, \eqref{Exchange},
5, \eqref{Syllogism}, Definition\ref{defn:daCosta(n)}, (MP)]

7 \ \ \ $(A^{\text{(n)}} \land B^{\text{(n)}}) \Rightarrow (A^n\land
B^n)$ \hfill[Definition\ref{defn:daCosta(n)}]

8 \ \ \ $(A^{\text{(n)}} \land B^{\text{(n)}}) \Rightarrow (A \ast
B)^n$ \hfill[6, 7, (2), (MP)]

\noindent This completes the proof.
\\$\square$\\
After all, we now know that systems $mZ_n$ do not form a hierarchy
but are equivalent to a single system which consists of IPC$^+$
together with axioms (9b), (10b)', (11b), (12b) and $mCZ_n$ can be
formulated by adding (13b) to these formulas. Note also that we
didn't make any use of axioms (10b)' and (11b) in proving Theorem
\ref{thm:axiom11redundant}.

Although it is not directly connected to the story of $mZ_n$ and
$mCZ_n$, it should be noted that propagation axiom for negation,
i.e. the following formula can be derived in an analogous manner:
\[ A^{\text{(n)}} \Rightarrow (\neg A)^{\text{(n)}} \]
Therefore, propagation axioms can be fully {\it proved} in systems
$mZ_n$ and $mCZ_n$.
\section{Semantics of negation based on Bikhoff's polarity}
In \cite{Majk08dc} (Proposition 3) was demonstrated that the
positive fragment of these two systems corresponds to the
distributive lattice $(X,\leq)$ (positive fragment of the Heyting
algebra), where the logic implication corresponds to the relative
pseudocomplement, $0,1$ are bottom and top elements in $X$ respectively.\\
Now we may introduce a hierarchy of negation operators
\cite{Majk06ml} for many-valued logics based on complete lattices of
truth values $(X,\leq)$, w.r.t their homomorphic properties: the
negation with the lowest requirements (antitonic) denominated
"general" negation can be defined in any complete lattice (see
example in \cite{Majk08dc}):
\begin{definition}
\textsc{Hierarchy of Negation operators:}  \label{def:negation} Let
$(X, \leq, \wedge, \vee)$ be a complete lattice. Then we define
the following hierarchy of negation operators on it:\\
1. A \verb"general" negation is a monotone mapping between posets ($\leq^{OP}$ is inverse of $\leq$),\\
$\neg:(X,\leq) \rightarrow (X, \leq)^{OP}$, such that $\{1 \}
\subseteq \{ y = \neg x~|~ x \in X \}$.\\
2. A \verb"split" negation is a general negation extended into
join-semilattice homomorphism,\\
$\neg:(X,\leq, \vee,0) \rightarrow (X, \leq, \vee,0)^{OP}$, with
$(X, \leq, \vee,0)^{OP} = (X, \leq^{OP}, \vee^{OP},0^{OP})$,
$\vee^{OP} =
\wedge$, $0^{OP} = 1$.\\
 3. A \verb"constructive" negation is a general negation
extended into full
 lattice homomorphism,
$~~ \neg:(X,\leq, \wedge, \vee) \rightarrow (X, \leq, \wedge,
\vee)^{OP}$,\\ with $(X, \leq, \wedge, \vee)^{OP} = (X, \leq^{OP},
\wedge^{OP}, \vee^{OP}) $, and $\wedge^{OP} = \vee$.\\
4. A \verb"De Morgan" negation is a constructive negation when the
 lattice homomorphism is an involution ($\neg \neg x = x$).
 \end{definition}
 The names given to these different kinds of negations follow from
 the fact that a split negation introduces the second right adjoint
 negation, that a constructive negation satisfies the constructive
 requirement (as in Heyting algebras) $\neg \neg x \geq x$,
 while De Morgan negation satisfies well known De Morgan laws:
 \begin{lemma}
 \textsc{Negation properties}): \label{lemma:negation}
 $~~ $ Let $(X, \leq)$ be a complete lattice. Then the following properties for negation operators hold: for any $x,y \in X$,\\
 1. for general negation: $~~ \neg(x \vee y) \leq \neg x \wedge \neg
 y$, $~~\neg(x \wedge y) \geq \neg x \vee \neg
 y$,\\ with $\neg 0 = 1$
 .\\
 2. for split negation: $~~ \neg(x \vee y) = \neg x \wedge \neg
 y$, $\neg(x \wedge y) \geq \neg x \vee \neg y$. It is an additive
 modal operator with right adjoint (multiplicative) negation $\sim:(X, \leq)^{OP}
 \rightarrow (X,\leq)$, and Galois connection $~~\neg x \leq^{OP}
 y~~$ iff $~~x \leq \sim y$, such that $~~x \leq \sim \neg x$ and
 $~~x \leq \neg \sim x $.\\
 3. for constructive negation: $~~ \neg(x \vee y) = \neg x \wedge \neg
 y$, $\neg(x \wedge y) = \neg x \vee \neg y$. It is a selfadjoint
 operator, $\neg = \sim$, with $x \leq \neg \neg x $ satisfying
 \verb"proto" De Morgan inequalities $~~\neg(\neg x \vee \neg y)
 \geq x \wedge y$ and $~~\neg(\neg x \wedge \neg y)
 \geq x \vee y$.\\
 4. for De Morgan negation ($\neg \neg x = x$): it satisfies also De Morgan laws $~~\neg(\neg x \vee \neg y)
 = x \wedge y$ and $~~\neg(\neg x \wedge \neg y)
 = x \vee y$, and is contrapositive, i.e., $~~x \leq y~~$ iff $~~\neg x \geq \neg
 y$.
\end{lemma}
Proof can be found in \cite{Majk06ml}.\\
\textbf{Remark:} We can see (as demonstrated in \cite{Majk08dc} that
the system $mZ_n$ without axiom (12b) corresponds to a particular
case of \emph{general} negation, that the whole system $mZ_n$
corresponds to a particular case of \emph{split} negation, while the
system $mCZ_n$ corresponds to a
particular case of \emph{constructive} negation.\\
The Galois connections can be obtained from  any binary relation
based on a set $\W$ \cite{Birkh40} (Birkhoff \emph{polarity})
in a canonical way:\\
If $(\W, \R )$ is a set with a particular relation based on a set
$\W$, $\R \subseteq \W \times \W$,  with mappings
$\lambda:\P(\W)\rightarrow \P(\W)^{OP}, \varrho:\P(\W)^{OP}\rightarrow \P(\W)$, such that for subsets $U,V \in \P(\W)$, \\
$\lambda U = \{w \in \W~|~\forall u \in U.( (u,w) \in \R) \}~~,
~~~~\rho V = \{w \in \W~|~\forall v \in V. ((w,v) \in \R )\}$,\\
where $(\P(\W), \subseteq)$ is the  \emph{powerset poset} complete
distributive lattice with bottom element empty set $\emptyset$ and
top element $\W$, and $\P(\W)^{OP}$ its dual (with  $\subseteq^{OP}$
inverse of $\subseteq$), then we have the induced Galois connection
$\lambda \dashv \rho$, i.e., $\lambda U \subseteq^{OP} V~$ iff
$~U \subseteq \rho V$.\\
It is easy to verify that $\lambda$ and $\rho$ are two antitonic
set-based operators which invert empty set $\emptyset$ into $\W$,
thus can be used as set-based negation operators. The negation as
modal operator has a long history \cite{Dose86}.\\
We denote by $~\mathfrak{R}~$ the class of such binary
incompatibility
relations $\R \subseteq \W \times \W$ which are also \verb"hereditary", that is\\
  if $~~(u,w) \in \R~$ and $~(u,w) \preceq (u',w')~~$ then $~~(u',w') \in
 \R$,\\
where $~(u,w) \preceq (u',w')~~$ iff $~~u \leq u'~$ and $~w \leq  w'$.\\
 Analogously to demonstration given in \cite{Majk08dc}, it is easy to see that, for any given  hereditary incompatibility
relation $\R $,  the additive algebraic operator $\lambda $ can be
used as the split negation for $mZ_n$ (or constructive negation,
when $\lambda$ is selfadjoint, i.e., $\lambda = \rho$, for $mCZ_n$).
\begin{coro} Each split negation (modal negation), based on the hereditary incompatible
relation of Birkhoff polarity, satisfies the Da Costa weakening
axioms (11) and (12).
\end{coro}
\textbf{Proof:} for the Birkhoff polarity we have that
 for any $U,V \subseteq \W$ holds the following additivity property,\\
  $~\lambda (U \bigcup V) = \lambda U \bigcup^{OP} \lambda V = \lambda U \bigcap \lambda
 V$, with $\lambda \emptyset = \emptyset^{OP} = \W$.\\
 It is well known that Heyting algebra operators ar closed for
 hereditary subsets, so that $\lambda$ applied to a hereditary
 subset $U$ has to result in a hereditary subset $\lambda(U)$ as
 well, and the Lemma 2 in \cite{Majk08dc} demonstrates that it is satisfied if the
 relation $\R$ is hereditary.\\
  It is enough now to prove that in $mZ_n$ the following formulae are valid (the logic negation operator $\neg$
  corresponds to the algebraic operator $\lambda$):\\
 $\neg(A \vee B) \equiv (\neg A \wedge \neg B)$, and $\neg 0 \equiv
 1$.\\
 Indeed, we can derive this as follows:
Indeed, we can derive this as follows:

1 \ \ \ $(1 \Rightarrow \neg 0) \Rightarrow ((\neg 0 \Rightarrow 1)
\Rightarrow ((0 \Rightarrow \neg 1) \wedge (\neg 1 \Rightarrow 0)))$
\hfill[(3)]

2 \ \ \ $(\neg 0 \Rightarrow 1) \Rightarrow ((0 \Rightarrow \neg 1)
\wedge (\neg 1 \Rightarrow 0))$ \hfill[1, (10b),(MP)]

3 \ \ \ $(0 \Rightarrow \neg 1) \wedge (\neg 1 \Rightarrow 0)$
\hfill[2, (11b), (MP)]

4 \ \ \ $\neg 0 \equiv
 1$ \hfill[3, by def. of $\equiv$],

 and,

 1 \ \ \ $(A\Rightarrow (A\vee  B)) \Rightarrow ( \neg(A \vee B) \Rightarrow \neg A)$
\hfill[(9b)]

2 \ \ \ $(B\Rightarrow (A\vee  B)) \Rightarrow ( \neg(A \vee B)
\Rightarrow \neg B)$ \hfill[(9b)]

3 \ \ \ $\neg(A \vee B) \Rightarrow \neg A$ \hfill[1,(6),(MP)]

4 \ \ \ $\neg(A \vee B) \Rightarrow \neg B$ \hfill[2,(7),(MP)]

5 \ \ \ $(\neg(A \vee B) \Rightarrow \neg A)\wedge (\neg(A \vee B)
\Rightarrow \neg B)$ \hfill[3,4, (3),(MP)]

6 \ \ \ $((\neg(A \vee B) \Rightarrow \neg A)\wedge (\neg(A \vee B)
\Rightarrow \neg B)) \Rightarrow (\neg(A \vee B) \Rightarrow (\neg A
\wedge \neg B))$ \hfill[ \eqref{Composition}]

7 \ \ \ $\neg(A \vee B) \Rightarrow (\neg A \wedge \neg B)$
\hfill[5, 6, (MP)]

8 \ \ \ $\neg(A \vee B) \equiv (\neg A \wedge \neg B)$ \hfill[7,
(12b), by def. of $\equiv$]

This completes the proof.
\\$\square$\\
This property holds for the constructive negation as well, thus for
the systems $mCZ_n$.\\
Thus, for these two paraconsistent systems we can define the Kripke
semantics in the similar way as for the intuitionistic logic.
\section{Conclusion}
In this paper we have slightly modified a weakening of negation
originally presented in the system $Z_n$ \cite{Majk08dc} in order to
obtain a paraconsistent logic, by eliminating the axiom $\neg 1
\Rightarrow 0$. This modified system $mZ_n$ has a split
negation.\\
Moreover if we preserve also the multiplicative property for this
weak  split negation we obtain the modified system $mCZ_n$ with a
\emph{constructive} paraconsistent negation which satisfies also the
contraposition law for negation.\\
Both systems have the negation that is different from the
(nonparaconsistent) intuitionistic negation (its algebraic
counterpart is different from the pseudocomplement of Heyting
algebras).  In both of them  the  the formula NEFQ is still
derivable, but  it does not hold the
falso quodlibet proof rule. Thus, they satisfy all da Costa conditions (from NdC1 to NdC4).\\
The Kripke-style semantics for these two paraconsistent negations
are defined as \emph{modal} negations: they are a conservative
extension of the positive fragment of Kripke semantics for
intuitionistic propositional logic \cite{Majk08dc}, where only the
satisfaction for negation operator is changed by adopting an
incompatibility accessibility relation for this modal operator which
comes from
Birkhoff polarity theory based on a Galois connection for negation operator.\\
If we denote by $Z_n^{-}$ the system obtained from $mZ_n$ by
eliminating the axiom (12b) (thus with the \emph{general} negation
in Definition \ref{def:negation}, that is \emph{only antitonic}),
then the da Costa axiom (12) can not be derived from the another
axioms (but the axiom (11) is still derivable from the antitonic
property of the negation). But in such a case, when we really need
the da Costa axiom (12), we are not able to define a Kripke-style
semantics for this negation operator, based on the Birkhoff
polarity. Consequently, this case needs more future investigations.

\textbf{Acknowledgments:} The author wishes to thank the graduate
student Hitoshi Omori (Graduate School of Decision Science and
Technology, Tokyo Institute of Technology, Japan) for his useful
comments and investigations of the properties of these modified
systems $mZ_n$ and $mCZ_n$. In particulary, he presented me the
formal proofs of the part of Lemma 1 and the proof of Theorem 2.


\bibliographystyle{IEEEbib}
\bibliography{mydb}

\end{document}

%% file: macros-general1.tex
\begingroup
\catcode`\~=11
\gdef\urltilde{\lower 0.6ex\hbox{~}}
\endgroup

 \renewcommand{\P}{\mathcal{P}}
 \newcommand{\R}{\mathcal{R}}

\newcommand{\W}{\mathcal{W}}